\documentclass[twocolumn,letterpaper,prb]{revtex4}
\usepackage{graphicx}
\usepackage{dcolumn}
\usepackage{amsmath}
\newcommand{\si}[1]{\ensuremath{_{\textrm{\scriptsize{#1}}}}}
\newcommand{\ssi}[1]{\ensuremath{_{\textrm{\tiny{#1}}}}}
\newcommand{\len}{\ell}
\newcommand{\vect}[1]{\boldsymbol{#1}}
\newcommand{\density}{\ensuremath{\rho\si{b}}}
\newcommand{\pt}[2]{\ensuremath{#1 \times 10^{#2}}}

\begin{document}

\title{Energetics and Possible formation and decay mechanisms of
  Vortices in Helium Nanodroplets}

\author{Kevin K. Lehmann}
  \email{lehmann@princeton.edu}
\author{Roman Schmied}
\affiliation{Department of Chemistry, Princeton University, Princeton,
  New Jersey 08544}

\date{\today}

\begin{abstract}
  The energy and angular momentum of both straight and curved vortex
  states of a helium nanodroplet are examined as a function of droplet
  size.  For droplets in the size range of many experiments, it is
  found that during the pickup of heavy solutes, a significant
  fraction of events deposit sufficient energy and angular momentum to
  form a straight vortex line. Curved vortex lines exist down to
  nearly zero angular momentum and energy, and thus could in principle
  form in almost any collision.  Further, the coalescence of smaller
  droplets during the cooling by expansion could also deposit
  sufficient angular momentum to form vortex lines.  Despite their
  high energy, most vortices are predicted to be stable at the final
  temperature (0.38\,K) of helium nanodroplets due to lack of decay
  channels that conserve both energy and angular momentum.
\end{abstract}

\pacs{PACS number}

\maketitle

Vortices are an almost unavoidable presence in bulk superfluid helium.
There is a rich history of studies of their properties and
interactions.\cite{Donnelly} It is natural to consider their possible
presence in the finite superfluid found in $^4$He
nanodroplets.\cite{Close98} The study of helium nanodroplets has been
quite active in recent years, starting with the spectroscopy of
embedded molecules as a probe for properties of nanodroplets, which
was introduced by Scoles and coworkers.\cite{Goyal92} However, despite
a now large body of work, no unambiguous signature of the presence of
vortices has yet been reported.  This is perhaps surprising, as a
vortex in a nanodroplet is expected to bind a molecular
impurity\cite{Dalfovo00, Draeger_thesis} and likely introduce a highly
anisotropic interaction potential, though no explicit calculation of
the magnitude of such an anisotropy has been reported to date.  The
vortex-induced anisotropy in molecular orientation, if large compared
to the rotational constant of the molecule in liquid helium, will
quench the molecular rotation and collapse the rotational structure
that is one of the hallmarks of molecular spectroscopy in helium
nanodroplets.\cite{Hartmann95,Callegari01}

Several calculations have been published for the energy of a droplet
with a straight vortex, with and without an atom or cylindrically
symmetric molecule aligned with the
vortex.\cite{Dalfovo00,Pi00,Mayol01,Draeger_thesis} As in bulk
superfluid, these calculations have found the energy of droplets with
a straight vortex to be significantly higher than that of vortex-free
droplets.\cite{Dalfovo00,Draeger_thesis} This result, combined with
the failure to observe vortex lines to date, have led some to propose
that vortices are unstable in helium nanodroplets and perhaps are
rapidly expelled.  In this paper, we revisit the energetics of vortex
lines, considering both the straight vortex line down the center of
the droplet that has previously been considered and curved vortex
lines that will rotate around the droplet due to their own flow field.
Furthermore, we consider the effect of angular momentum conservation
on the possible formation and decay of vortices in droplets.  It turns
out that the linear and curved vortices contain considerably less
excitation energy per unit angular momentum than the final states
accessible by decay through the mechanisms considered.  This suggests
that despite their higher energy than the ground state, droplets with
such vortices are the lowest possible states with high angular
momentum, and thus should in fact be stable to decay.

\section{Hollow Core Model of Vortex Lines}

There have been several microscopic calculations of the properties of
straight vortex lines in helium nanodroplets, both pure and doped with
atomic or molecular
solutes.\cite{Dalfovo00,Lekner00,Pi00,Mayol01,Barranco02,Draeger_thesis}
In this work, we exploit the phenomenological description known as the
hollow core model.  The numerous microscopic treatments of vortex
lines in two and three dimensions have largely confirmed its
qualitative applicability,\cite{Chester68,
  Dalfovo92,Ortiz95,Vitiello96,Sadd97,Giorgini96} and it is widely
used to describe the properties of vortex lines and rings in bulk
liquid helium.\cite{Donnelly,Rayfield64}

The vortex is surrounded by a circulating flow, characterized by an
irrotational velocity field ($\vect{\nabla} \times \vect{v} = 0$).  In
the simple case of a straight vortex in bulk helium, the magnitude $v$
of the flow velocity at any given point is inversely proportional to
the distance of that point from the center of the vortex, $ v = \hbar
/ m r $ ($m$ is the mass of a $^4$He atom).  This leads to a kinetic
energy density whose volume integral diverges as the vortex is
approached ($ r \rightarrow 0$).  In the hollow core model, the helium
number density $\rho$ is taken to be zero inside a cylinder of radius
$a$ and equal to the bulk value $\density = 0.0218$\,\AA$^{-3}$
outside of this cylinder.  Such a discontinuous change in density is
unphysical, but microcanonical calculations have confirmed a nearly
hollow core, though with a smooth transition of the density to the
bulk value.\cite{Dalfovo92, Vitiello96} It is noted that our model
treats the helium density on the outer boundary of the droplet as
abruptly going to zero, while it is known that in fact the surface of
liquid helium is diffuse, with a thickness of $\approx
6-8$\,\AA.\cite{Harms98b} A hollow core radius of $\tilde{a} =
1.00$\,\AA\ was found experimentally\cite{Rayfield64} to best
reproduce the measured energy and velocity of vortex rings in bulk
liquid helium; matching our expression of the velocity of vortex rings
to the expression in Ref.~\onlinecite{Rayfield64} requires us to use
$a = \tilde{a}/\sqrt{e} = 0.607$\,\AA\ (see
Appendix~\ref{sec:HCmodel}).  Any normal fluid component is neglected
as helium nanodroplets in this size range have no thermally excited
phonon excitations.\cite{Brink90}

First we consider a straight vortex through the center of a helium
nanodroplet.  Because of the loss of spherical symmetry, we will
consider that droplet to have the shape of an ellipsoid, with axial
radius $B$ and equatorial radius $A$; the density is assumed to be
uniform, equal to the bulk value \density, and dropping to zero at the
ellipsoidal surface.  In this geometry, the flow field is the same as
for the bulk straight vortex described above.  The kinetic energy,
$E\si{sv}$, of this straight vortex flow field is given by
\begin{multline}
\label{eq:Esv1}
E\si{sv} = \frac{h^2 \density B}{2 \pi m}
\left[ \frac{1}{2} \ln \left( \frac{A +\sqrt{A^2 - a^2}}
{A - \sqrt{A^2 - a^2}} \right)
- \frac{\sqrt{A^2 - a^2}}{A} \right]\\
\rightarrow \frac{h^2 \density B}{2 \pi m} \left[
\ln \left( \frac{2 A}{a} \right) - 1 \right]
\qquad \textrm{for}\ A\gg a.
\end{multline}
For a spherical droplet of $N$ atoms, $B = A = R = \sqrt{r_0^2 N^{2/3}
  + a^2}$, with $r_0 = \left(4\pi\density/3\right)^{-1/3} =
2.22$\,\AA.

An interesting question is the extent to which the large angular
momentum of the vortex will distort the otherwise spherical droplet.
Making the droplet oblate will reduce the length of the vortex line
and thus lower its energy.  However, this will also increase the
surface energy of the droplet.  The latter is the product of the
surface tension of bulk liquid helium,
$\sigma=0.272$\,K\,\AA$^{-2}$,\cite{Deville96} and the surface area
of an ellipsoid, $S = 2 \pi A^2 + \frac{\pi B^2}{e} \ln \left( \frac{
    1 + e}{1 - e} \right)$, with the eccentricity
$e=\sqrt{1-(B/A)^2}$.  The droplet distortion is found by minimizing
the sum of the vortex kinetic energy (given by Eq.~(\ref{eq:Esv1}))
and the surface energy of the droplet, at constant droplet volume $V =
\frac{4\pi}{3} \frac{B}{A} (A^2 - a^2)^{3/2}$.
Table~\ref{tb:droplets} gives the resulting eccentricity, $e$, and the
associated stabilization energy (reduction in energy from a spherical
droplet of the same volume with vortex).  It is seen that the
distortion from spherical symmetry is small, in the sense that $B/A =
\sqrt{1-e^2}$ is close to 1, despite the high angular momentum of the
vortex, and this distortion will be neglected in the rest of this
paper.

\begin{figure}[htbp]
  \includegraphics[width=3in]{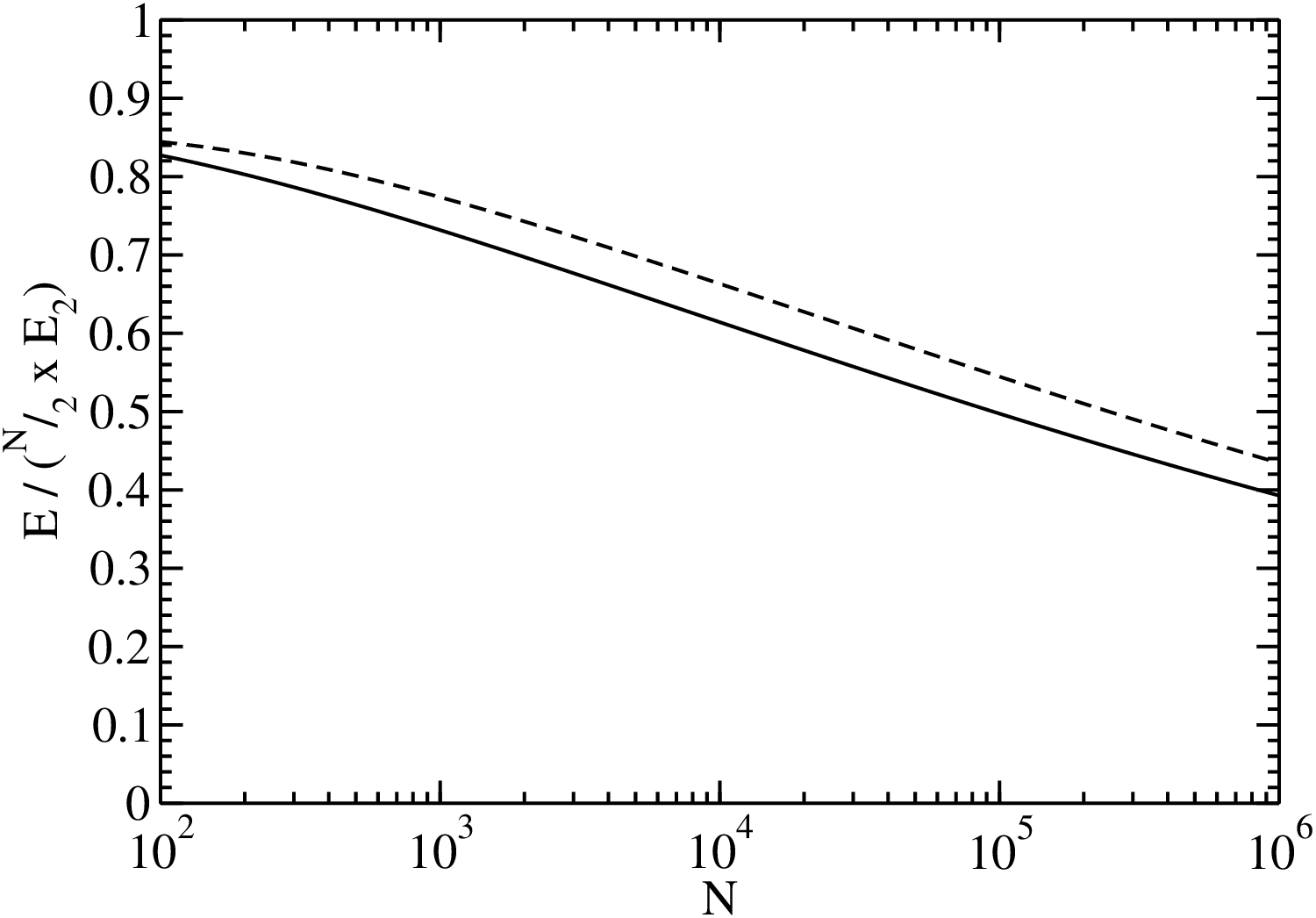}
  \caption{Energy of the straight vortex line solution as a function
    of the number of He atoms in the droplet, calculated both by the
    Hollow Core model (solid line, Eq.~(\ref{eq:Esv1})) and by finite
    range Density Functional Theory (dashed line,
    Ref.~\onlinecite{Dalfovo00}).  Energies are displayed as a
    fraction of the minimum energy required to deposit $N \hbar$ units
    of angular momentum into ripplon excitation modes.}
  \label{fig:Esv}
\end{figure}

Dalfovo \emph{et al.}\cite{Dalfovo00} used finite range Density
Functional Theory (DFT) to calculate the energy and core shape of a
straight line vortex for a range of droplet sizes ($N = 50 - 1000$).
While their core has a smooth density profile, they report a core
radius of the order of $1-2$\,\AA.  Figure~\ref{fig:Esv} shows a plot
of the vortex energy calculated by the hollow core model as a function
of the number of helium atoms in the droplet, along with the same
quantity estimated by the Density Functional\cite{Dalfovo00} method.
The hollow core model appears to slightly underestimate the vortex
energy with respect to DFT; agreement with the DFT calculation can be
made almost quantitative if a value of $a = 0.56$\,\AA\ is used.  The
core radius used in the hollow core model is to be interpreted as an
effective radius that reproduces the energetics and velocity of
experimental vortices, and not more than an estimate of the real core
radius.
 
The angular momentum associated with a straight vortex is $N \hbar$,
\emph{i.e.}, one unit per helium atom.  Both methods agree that the
vortex energy per unit angular momentum is lower than for any other
excitation mode of a pure helium droplet:\cite{Brink90}
Figure~\ref{fig:Esv} shows the vortex excitation energies relative to
the energy required to produce $N/2$ quanta of $L = 2$ ripplons, which
is the lowest energy state based upon quasiparticle excitations that
has the same total angular momentum as the straight vortex.
Table~\ref{tb:droplets} shows a comparison of the hollow core model
excitation energy of a straight vortex and this lowest ripplon, and
their energies per unit angular momentum for droplets of the size
range that span most helium nanodroplet experiments.  As will be
discussed below, it is the relatively low energy per unit angular
momentum that is key to the proposed metastability of vortex line
solutions.

\begin{table*}[htb]
\centering
\caption{Properties of Helium droplets and their straight vortex
  solutions as a function of the number of Helium atoms.
\label{tb:droplets}
\newline}
\begin{tabular}{lcccccl}
\hline\hline
He Number & $10^2$ & $10^3$ & $10^4$ & $10^5$ & $10^6$ \\
\hline
Radius $R$                  &  1.033 & 2.222 & 4.785 & 10.31 & 22.21  & nm \\
Helium Binding Energy         & 4.72  & 6.03  & 6.63 & 6.91 & 7.04 & K \\
Total Thermal Ripplon Energy      & 0.413 & 3.56 & 18.5 & 87.0 & 403. & K \\
$L = 2$ Ripplon Energy      & 1.05 & 0.332 & 0.105 & 0.0332 & 0.0105 & K \\
Thermal Ripplon $\sqrt{\left<L(L+1)\right>}$& 
1.54 & 8.54 & 38.7 & 173. & 781. & $\hbar$\\
Vortex Energy $E\ssi{DFT}$          &
44.3 & 129. & 349. & 905. & 2283. & K \\
Vortex Energy $E\ssi{HC}$ (Hollow Core) &
43.4 & 122. & 323. & 827. & 2064.  & K \\
Vortex Angular Momentum $L\ssi{v}$    &
$10^2$ & $10^3$ & $10^4$ & $10^5$ & $10^6$ & $\hbar$ \\
$E\ssi{HC}/L\ssi{v}$    & 56.8 & 15.9 & 4.23 & 1.08 &  0.270 & GHz \\
Lowest Ripplon $E/L$ & 68.6 & 21.8 & 6.88 & 2.18 & 0.688  & GHz \\
\hline
Vortex-induced eccentricity        & 0.53 & 0.44 & 0.35 & 0.27 & 0.20 & \\
B/A                                   & 0.85 & 0.90 & 0.94 & 0.96 & 0.98 &  \\
Deformation Stabilization Energy   & 2.00 & 3.74 & 6.16 & 9.22 & 12.9 & K \\
\hline
Max.\ $L$ loss by 1 atom evaporation & 26    & 96    & 345   & 1198   & 4090    & $\hbar$\\
Max.\ $L$ loss by $n$ atom evaporation & 41    & 222   & 1213   & 6561   & 35000   & $\hbar$\\
for evaporation of $n=$                & 5     & 10     &  25   & 59     & 146     & atoms\\
Max.\ $L$ loss by fission            &  28   & 210   & 1488  & 10150   & 67400    & $\hbar$\\
\hline
$x_0$ for stability limit of curved vortex  & 0.72  & 0.76 & 0.81 & 0.85  & 0.89 & $R(N)$
\\ 
Min.\ $L$ for stable vortex          &  18  &  131  &  850  & \pt{5.14}{3} & \pt{2.96}{4} &
$\hbar$ \\
Min.\ $E$ for stable vortex          &  9.42 & 27.2 & 62.1 & 126.  & 242. & K \\
Max.\ $v$ for stable vortex          &  56.3 & 39.0 & 27.6 & 19.5 & 13.8 & m s$^{-1}$\\
Max.\ kinetic energy of 100\,u dopant                 & 19.1 & 9.1 & 4.6 & 2.3 & 1.1 & K\\
\hline\hline
  \end{tabular}
\end{table*}

\section{Curved Vortex Line Solutions}

We now turn to more general vortex line solutions.  The flow field
around a general vortex line in bulk helium is homologous to that of
the magnetic field around a curved wire that follows the vortex line.
The ``current'' in this wire is proportional to the quantum of
circulation, $\kappa = h/m$.  Thus the flow field $\vect{v}(\vect{r})$
is given by the Biot-Savart equation\cite{Bauer95}
\begin{equation}
\label{eq:BS}
\vect{v}(\vect{r})
= \frac{\kappa}{4 \pi} \int\si{vortex}
\frac{ \left( \vect{s} - \vect{r} \right) \times 
\mathrm{d}\vect{s}}
{\left| \vect{s} - \vect{r} \right|^3}.
\end{equation}
The vortex must either form a closed loop or end at a boundary of the
superfluid helium. $\vect{v}(\vect{r})$ must not have a normal
component at any boundary of the superfluid.  This implies that the
vortex must intersect a helium boundary at normal incidence.

A curved vortex will move in its own flow field.  When the local
radius of curvature of the vortex $\mathcal{R}(\vect{s})$ is large
compared to its core radius $a$, the motion of the vortex core can be
calculated using the Local Induction Approximation (see
Appendix~\ref{sec:shape} and Ref.~\onlinecite{Rayfield64}).  The shape of
the vortex core surface is determined by the condition of no helium
flow across it, and turns out to be circular in cross section as long
as $a\ll \mathcal{R}$.  For finite core size, matching this boundary
condition requires either that the core shape be altered or that one
add an irrotational solution to Laplace's equation that corrects the
normal component of velocity.  It is not evident to the authors which
change to make, and so these errors are neglected in the rest of the
paper.

We now specialize to the case of a vortex line inside a spherical
droplet of radius $R$.  Muirhead, Vinen, and
Donnelly\cite{Muirhead84} showed that for any arbitrary vortex shape
in a spherical droplet, the boundary conditions of the flow velocity
on the surface of the droplet can be satisfied by continuing an image
vortex outside the droplet.  Each point on the vortex $\vect{s}$ ($s$
is the magnitude of this vector) generates an image point
$\vect{s}\si{i} = (R/s)^2 \vect{s}$ with vorticity equal to $-\kappa
s/R$.  Vorticity is conserved by attaching to each point along the
image vortex a radially pointing vortex going to infinity that has a
circulation strength given by the decrease in circulation of the image
along its length.  These all combine in the Biot-Savart equation to
give
\begin{multline}
\label{eq:vs2}
\vect{v}(\vect{r}) =
\frac{\kappa}{4 \pi} \int\si{vortex} \left(
\frac{ (\vect{s}-\vect{r}) \times \mathrm{d}\vect{s}}
{|\vect{s}-\vect{r}|^3}
-\frac{s}{R} \cdot \frac{ (\vect{s}\si{i}-\vect{r}) \times
\mathrm{d}\vect{s}\si{i}}
{|\vect{s}\si{i}-\vect{r}|^3}
 \right. \\
\left.
- \frac{\vect{s}\cdot \mathrm{d}\vect{s}}{R}
\cdot
\frac{\vect{r}\times \vect{s}}
{s^2 |\vect{s}\si{i}-\vect{r}|^2
+ s|\vect{s}\si{i}-\vect{r}|
(R^2-\vect{r} \cdot \vect{s})}\right),
\end{multline}
where
\begin{equation}
\label{eq:dsi}
\mathrm{d}\vect{s}\si{i} = R^2 \left(
\frac{\mathrm{d}\vect{s}}{s^2}
- \frac{2 ( \vect{s} \cdot \mathrm{d}\vect{s})
\vect{s}}{s^4}  \right).
\end{equation}
The three terms in the integral arise from the vortex, image vortex,
and vorticity conserving radial vortex lines, respectively, and make
contributions that decrease in magnitude the order given.

\begin{figure}[htbp]
\includegraphics[width=3in]{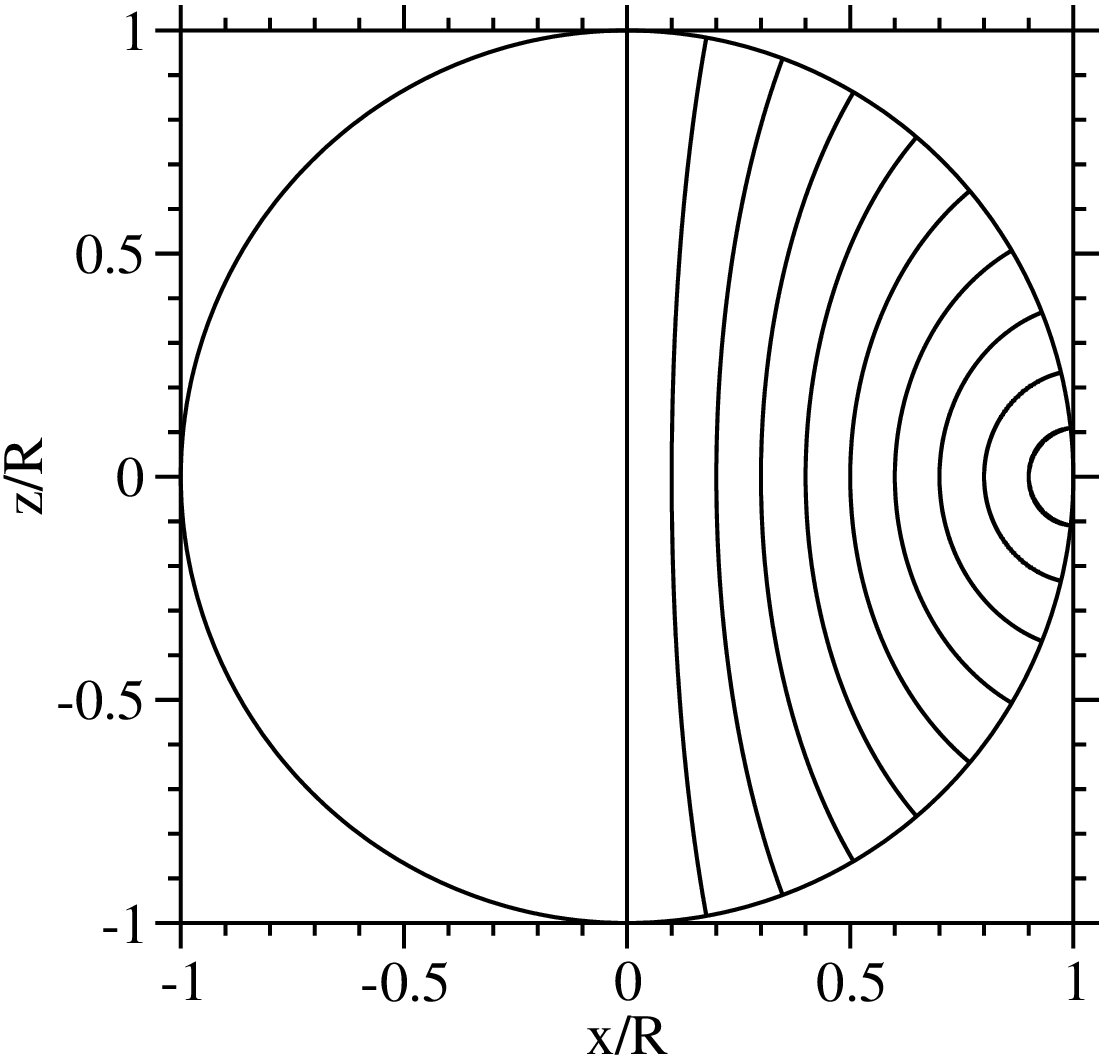}
  \caption{Curved vortex solutions for different values of $x_0$, the distance
    of closest approach to the $z$ axis, in a droplet of $N=10000$ He
    atoms.}
  \label{fig:vortex_lines}
\end{figure}

We seek curved vortex line solutions $\{x(\len),z(\len)\}$ that rotate
at constant angular velocity $\Omega$ around the $z$ axis, which
implies that these vortex line solutions will have constant shape.
Appendix~\ref{sec:shape} gives the numerical procedure used to
determine these solutions.  Figure~\ref{fig:vortex_lines} shows some
of the solutions for several values of $x_0$, the distance of minimum
approach to the $z$ axis.

Bauer, Donnelly, and Vinen\cite{Bauer95} showed that the total
angular momentum $\vect{L}\si{v}$ and kinetic energy $E\si{v}$ of the
helium flow (which are defined by volume integrals for the
corresponding densities) can be reduced to two surface integrals
\begin{align}
\label{eq:L}
\vect{L}\si{v} & = m \density \kappa \int \vect{r} \times
\mathrm{d}\vect{S}\\
\label{eq:Ev}
E\si{v} & = \frac{1}{2} m \density \kappa \int \vect{v}
\cdot \mathrm{d}\vect{S},
\end{align}
where the integration is over the the region in the $x z$ plane
bounded by the vortex and the surface of the droplet. $\vect{v}$ in
Eq.~(\ref{eq:Ev}) is given by Eq.~(\ref{eq:vs2}).  The origin for
vector $\vect{r}$ in Eq.~(\ref{eq:L}) must be taken as the center of
the sphere so that the outer surface of the sphere does not contribute
to this integral expression.  These expressions have neglected a
contribution of the integral over the surface of the vortex core, but
that should be small as long as $\mathcal{R} \gg a$ as required by our
approximations. In the Appendix, we give explicit expressions for the
lowest order (in $a/\mathcal{R}$) core surface corrections to
$L\si{v}$ and $E\si{v}$, which are the calculations reported below.
For the vortices considered here, $\vect{L}\si{v}$ is parallel to the
$z$ axis owing to the symmetry of the vortex and its flow field with
respect to reflection in the $x y$ plane.  Calculation of $E\si{v}$
requires evaluation of a triple integral, with an integrand that is
nearly singular along one of the edges of the integration domain; one
of the integrations can be done analytically, leaving a double
numerical integration. In the same way, the double integral for
$L\si{v}$ can be reduced to a single numerical integration.

The treatment as yet is for vortex lines as classical objects with
angular momentum pointing in a definite direction in space.  By the
standard rules of semiclassical quantization, vortex eigenstates can
be constructed as linear combinations of the vortex lines with angular
momentum pointing in all possible directions, with an amplitude for
each direction given by a spherical harmonic.  Such states are
eigenstates of both the total helium angular momentum (quantum number
$L\si{v}$) and also its projection on the laboratory $Z$ axis (quantum
number $M\si{v}$).  The vortex lines we have discussed have total
squared angular momentum that span the range $(N \hbar)^2$ to 0 as
$x_0$ goes from $0$ to $R$.  Semiclassical quantization will restrict
$x_0$ to $N-1$ values with total angular momentum quantum number equal
to $L\si{v}=N-1, N-2, \ldots, 1$.  The straight line vortex is not
allowed since the length of the total angular momentum of the allowed
vortex solutions is $|\vect{L}\si{v}| = \hbar
\sqrt{L\si{v}(L\si{v}+1)}$, and the straight vortex solution has
$|\vect{L}\si{v}| = \hbar N$.  Each vortex solution with integer
$L\si{v}$ quantum number has a ($2L\si{v}+1$)-fold $M$-degeneracy.
This set of solutions gives $N^2$ semiclassical vortex states for a
droplet of $N$ helium atoms.  In principle, there are other states
that involve vibrational excitation of the vortex
lines\cite{Kiknadze02} around these Bauer--Donnelly--Vinen vortex
solutions, but these will not be considered in this work.

\begin{figure}[htbp]
  \includegraphics[width=3in]{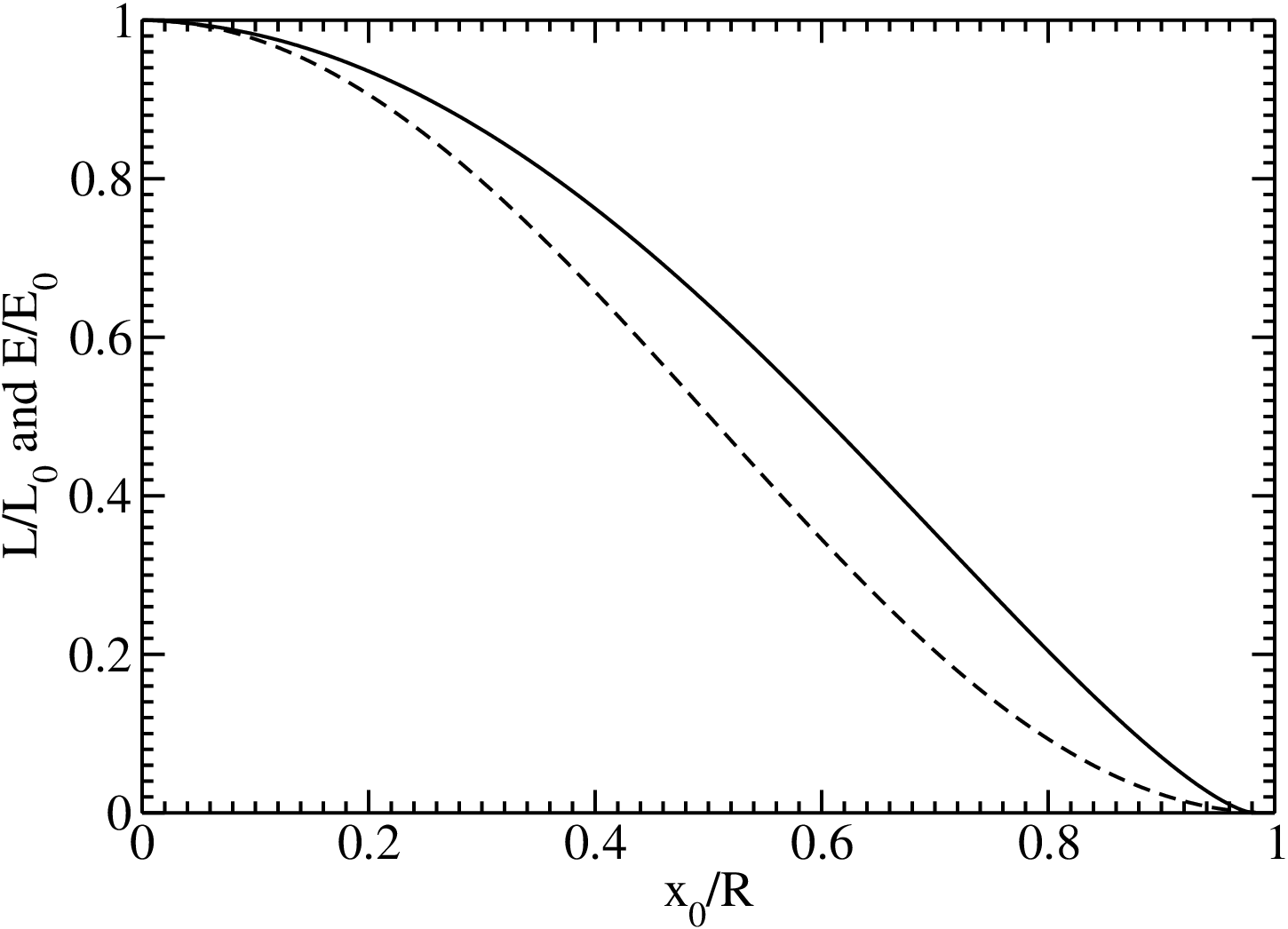}
  \caption{Energy (solid line) and angular momentum (dashed line)
    of a curved vortex solution as a function of the distance of
    closest approach of the vortex from the $z$ axis.  These results
    are for a $N = 10^4$ helium atom droplet. The energy $E_0$ of the
    straight vortex is given in Expression~(\ref{eq:Esv1}); its
    angular momentum is $L_0=N \hbar$.}
  \label{fig:ELvsS0}
\end{figure}

\begin{figure}[htbp]
  \includegraphics[width=3in]{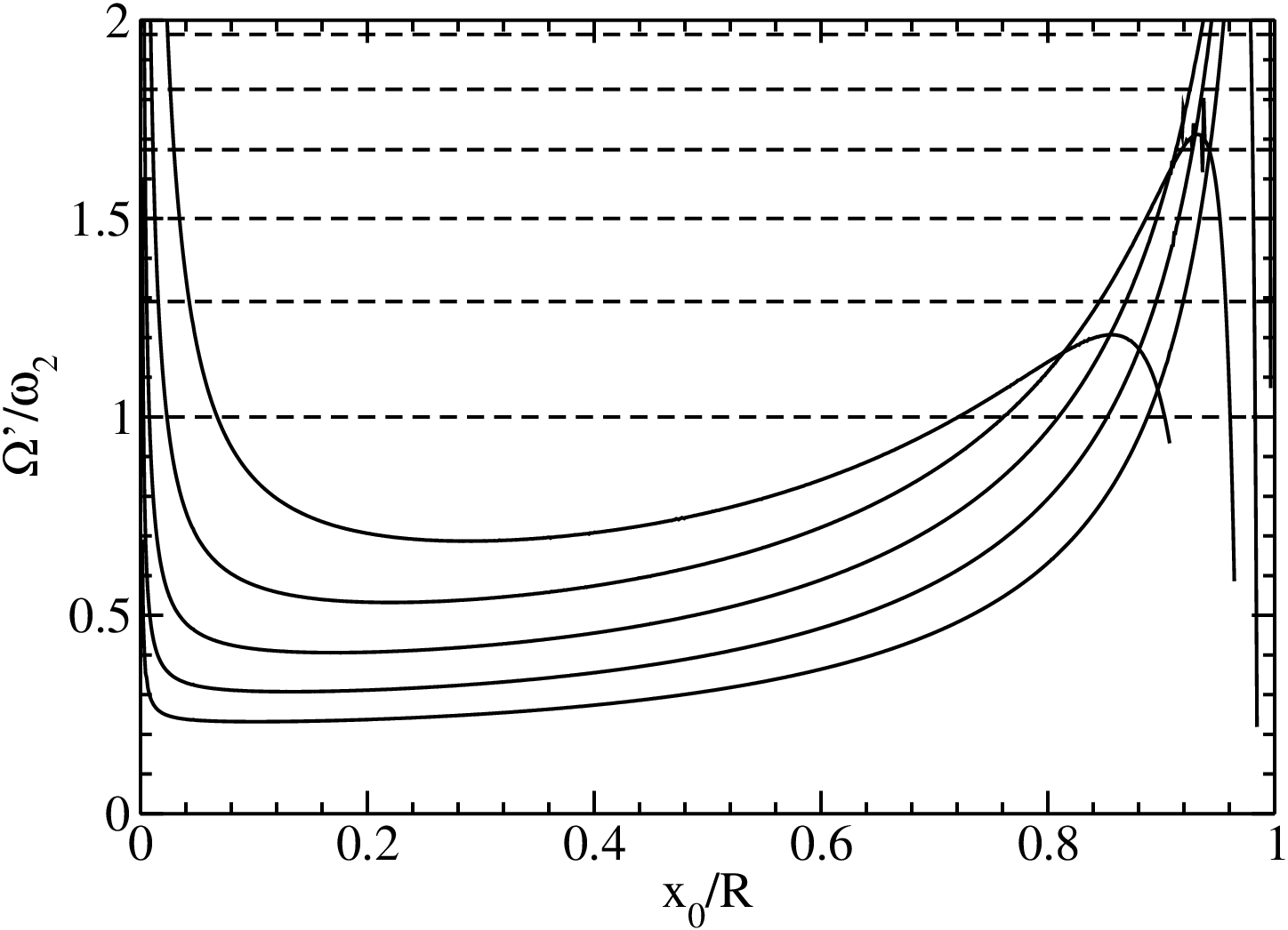}
  \caption{Comparison of the vortex ``angular velocity''
    $\Omega'=\mathrm{d}E/\mathrm{d}L$ and the angular velocities
    $\omega_L$ of the lowest ripplons ($L=2\ldots 7$), as a function
    of distance of closest approach to the $z$ axis, for droplets with
    $N = 10^2,10^3,10^4,10^5,10^6,10^7$ (top to bottom) He atoms. All
    angular velocities are shown as ratios to $\omega_2=\sqrt{8\pi
      \sigma/(3m N)}$.}
  \label{fig:OmegaRipplon}
\end{figure}

Figure~\ref{fig:ELvsS0} shows the calculated vortex energy and angular
momentum for a droplet with $N = 10^4$ helium atoms.  It is seen from
this figure that the energy of the vortex drops as the vortex is moved
off axis, going monotonically to zero as the vortex is ``pushed out''
of the droplet.  By energetic considerations alone, this would imply
that the vortex solutions are unstable.  However, for an isolated
droplet, one must also conserve angular momentum.  A vortex can lower
its energy by producing a ripplon if the derivative of the vortex
energy with respect to total angular momentum, $\Omega'$ (with units
of angular velocity), is greater than the $E/L$ ratio of the $L = 2$
ripplon.  Figure~\ref{fig:OmegaRipplon} shows $\Omega'$ for vortices
as a function of their minimum approach distance to the $z$ axis,
normalized to the energy per unit angular momentum of this ripplon
mode.  $\Omega'$ was evaluated by calculation of both $E$ and $L$ for
$10^3$ values of $x_0$ for each droplet size, and using $\Omega' =
\frac{dE}{dx_0} / \frac{dL}{dx_0}$, with each derivative evaluated by
finite difference of the calculated points.  It is evident that the
vortex is stable to ripplon production (has a normalized value of
$\Omega'$ less than unity) for most of its range.
Table~\ref{tb:droplets} list the minimum energy and angular momentum
of the curved vortex states that are thus stable.

\begin{figure}[htbp]
  \includegraphics[width=3in]{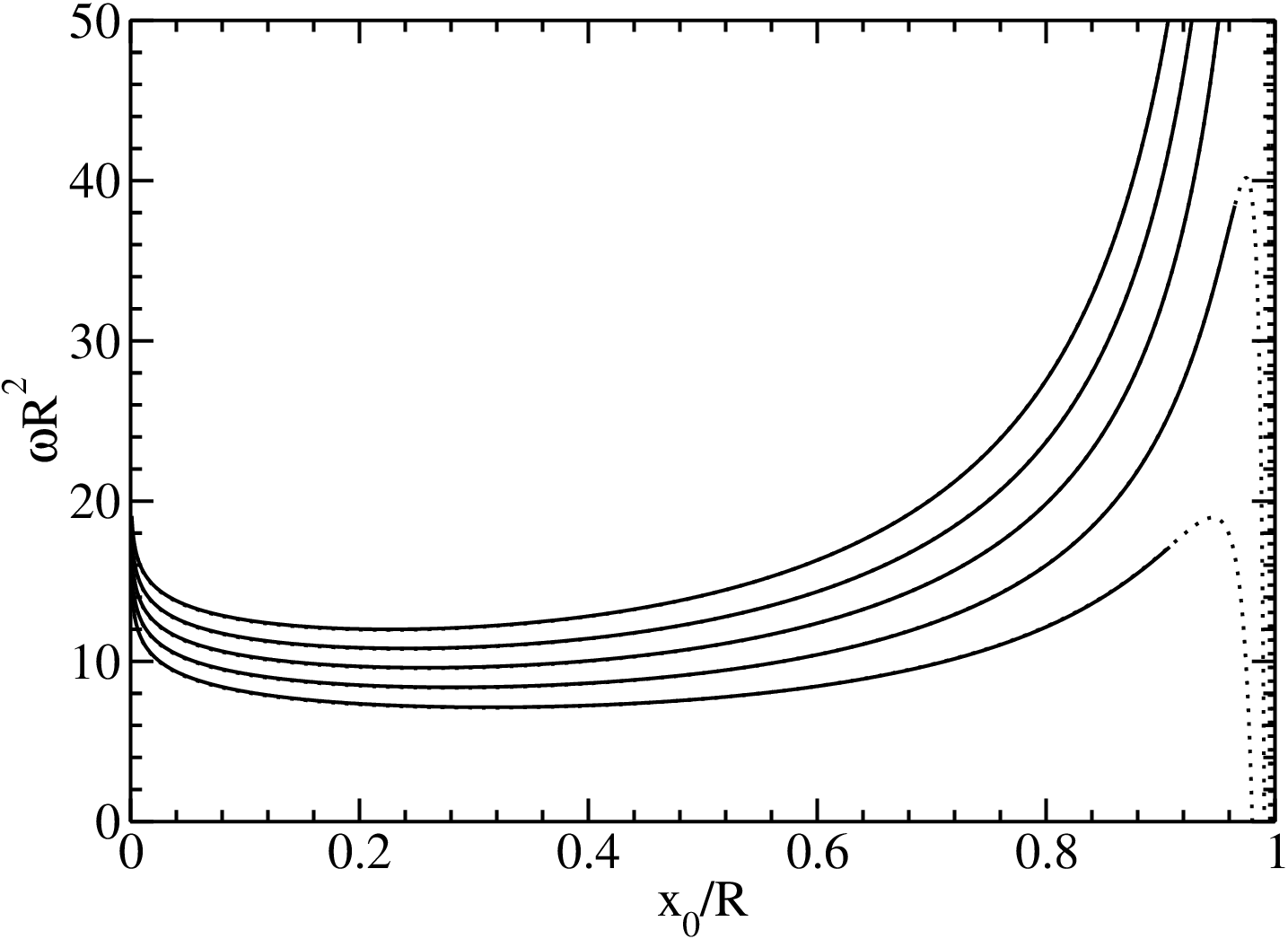}
  \caption{Angular velocity as a function of distance of closest approach,
    calculated for droplets with $N = 10^2$, $10^3$, $10^4$, $10^5$,
    $10^6$ (bottom to top) He atoms. Plotted is $\omega R^2 = 4\pi
    \Omega R^2/\kappa$, a dimensionless quantity. Solid lines are
    computations using the model described in
    Appendix~\ref{sec:shape}; dotted lines use the model of
    Ref.~\onlinecite{Bauer95}. As discussed in the Appendix, the
    local-induction approximation breaks down as $x_0\to R$.}
  \label{fig:Omega}
\end{figure}

\begin{figure}[htbp]
  \includegraphics[width=3in]{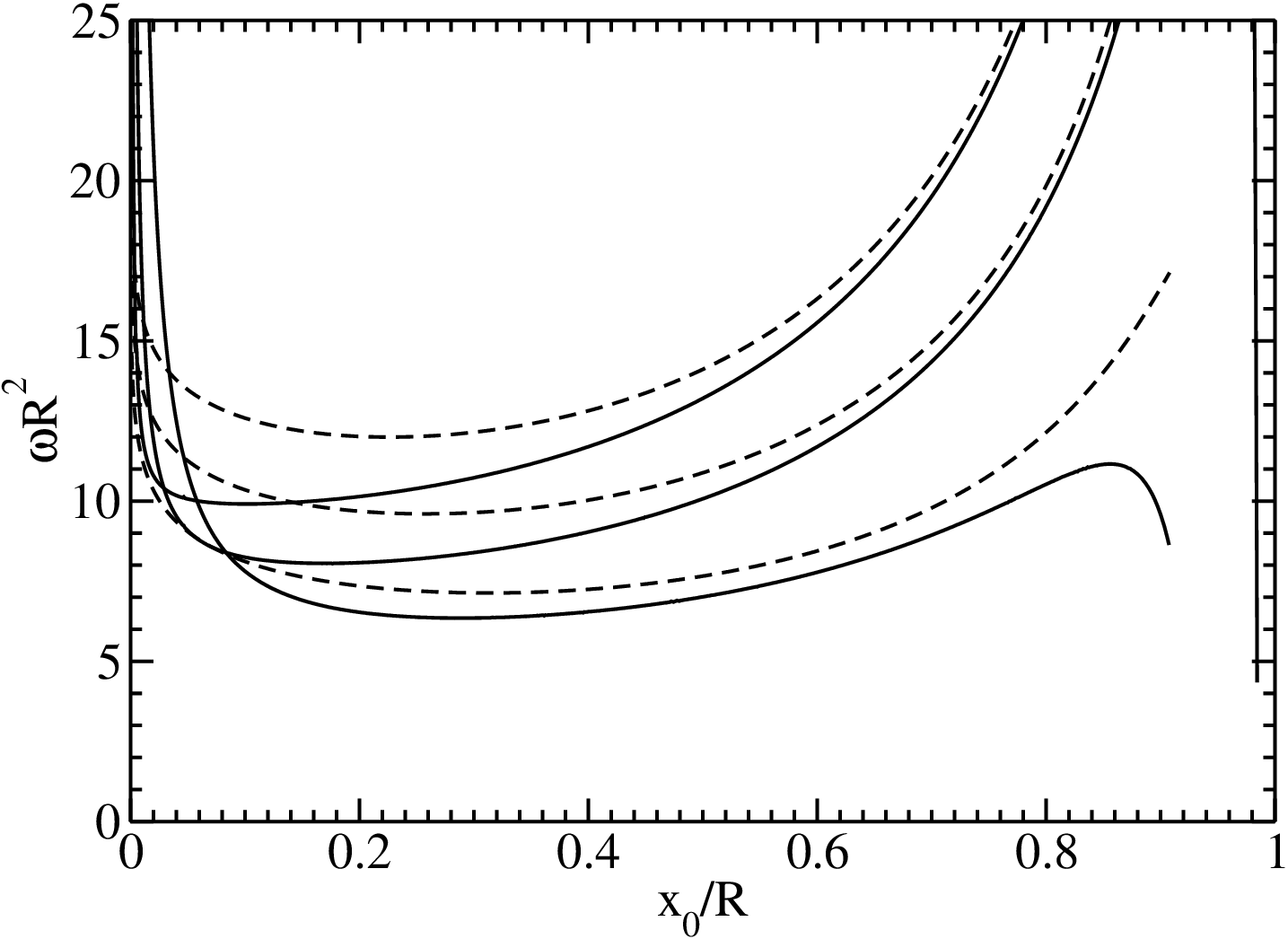}
  \caption{Comparison of the computed reduced angular velocity
    $\omega = \frac{4\pi}{\kappa} \Omega$ (dashed lines) and the
    quantity $\omega'=\frac{4\pi}{\kappa} \mathrm{d}E/\mathrm{d}L$
    (solid lines), for droplets with $N=10^2,10^4,10^6$ atoms (bottom
    to top).}
  \label{fig:Omega_comp}
\end{figure}

Because the energy and angular momentum in the hollow core model arise
entirely from helium motion, we had anticipated that $\Omega'$ would
be equal to $\Omega$, the angular velocity of the curved vortex
solution around the $z$ axis.  For vortex rings in the local induction
approximation, Rayfield and Reif\cite{Rayfield64} found that the
vortex velocity satisfied $v = dE/dp$, where $p$ is the net helium
linear momentum due to helium flow around the vortex ring.  This
insures that as $x_0 \rightarrow R$ (but $R - x_0 \gg a$ so the local
induction approximation still holds), $\Omega' \rightarrow \Omega$.
For a vertical vortex in a cylinder (where $E$ and $L$ have simple
analytical solutions), we have demonstrated that $\Omega' = \Omega'$.
For our solutions of curved vortex lines in a spherical droplet, we
however find that $\Omega' \ne \Omega'$, as demonstrated in
Figure~\ref{fig:Omega_comp}, where both are compared as a function of
$x_0/R$ for various droplet sizes.  It appears from our calculations
that $\Omega' \rightarrow \Omega$ as $N \rightarrow \infty$ (except
for the log divergence of $\Omega$ as $x_0 \rightarrow 0$), but the
rate of convergence is rather slow.  It remains to be established
whether the discrepancy between $\Omega$ and $\Omega'$ is due to an
error in calculation or whether it is real.
  
It is evident from figure~\ref{fig:Omega_comp} that $\Omega$ peaks for
the straight vortex ($x_0=0$); more careful analysis shows that there
is in fact a logarithmic divergence: $\Omega \approx
\frac{\kappa}{2\pi R^2} \ln \frac{4R^2}{a x_0 e}$ as $x_0 \rightarrow
0$.  In the limit that the vortex line moves near the surface of the
droplet, our solutions no longer provide meaningful estimates as the
vortex ring radius $R - x_0$ approaches the value of the core radius
$a$.  In addition to the expected breakdown of the local induction
approximation in this limit, such a vortex would be localized in the
region of highly inhomogeneous density near the helium surface.
Feynman has suggested that the limit of a vortex ring of atomic scale
is the roton quasiparticle excitation.\cite{Feynman-Stat-Mech}

In order to test that there was no numerical error in the evaluation
of the integral expressions for $L\si{v}$ and $E\si{v}$ reported in
this work, calculations for selected values of $a/R$ were performed
with totally independent routines written by the two authors, using
distinct programing packages (Mathematica and Mathcad), and the values
of $\Omega$, $L\si{v}$, and $E\si{v}$ were found to agree to six
significant figures.

\section{Possible Spectroscopic Signatures of Molecules bound to Vortices}

It is well know in bulk liquid helium that impurities tend to bind to
vortex lines or vortex rings.  The binding of ions to vortex rings was
the basis of the remarkable experiment of Rayfield and
Reif\cite{Rayfield64} that measured many of the properties of ring
vortices.  Calculations by both Dalfovo \emph{et al.}\cite{Dalfovo00}
and by Draeger and Ceperley\cite{Draeger_thesis} have demonstrated
that atomic or linear molecular impurities in a helium nanodroplet
bind to linear, maximum angular momentum vortex lines.  For
technical reasons, both these calculations required cylindrical
symmetry, and thus could not study the energy change upon rotation of
the molecule with respect to the vortex axis.  However, it is
generally believed that the binding energy is maximized by aligning
the linear molecular axis with the vortex since this displaces the
maximum amount of helium from near the vortex core, where it has the
highest kinetic energy; however, the increased helium density in the
first solvation layer around the solute may in part counteract this
effect.  This suggests that the anisotropy in the binding energy
should be of comparable magnitude as the binding itself, but this
natural expectation has yet to be checked by calculations with a
realistic helium density profile.  It seems likely that the magnitude
of the anisotropy of the binding for linear molecules, such as HCN and
HCCCN, is much higher than the rotational constant of these molecules
in liquid helium.  If this was correct, then the gas phase-like
rotational structure of these molecules would be quenched, leading to
a rovibrational spectrum dominated by a Q branch; such Q branches are,
however, absent in the observed spectra of these
molecules.\footnote{There have been weak Q branches observed in the
  spectra of polymers of these species, but these have recently been
  assigned to thermal excitations.}  This suggests that at most a
small fraction of the droplets probed in these experiments could have
such a linear vortex line excitation.  It is further noted that the
vibrational spectrum of the spherical top molecule SF$_6$ is also
expected to be a sensitive probe for the presence of a vortex line, as
the highly anisotropic helium density created by the vortex core would
lift the triple degeneracy of IR active fundamental modes, producing a
spectroscopic structure in qualitative disagreement with what has been
observed.\cite{Hartmann95}

Molecules are also expected to bind and align with curved vortex
lines, as long as the solute kinetic energy required for it to move
with the curved vortex line does not exceed the binding energy of the
solute molecule to the stationary vortex.  The velocity of a curved
vortex increases smoothly from zero when $x_0 = 0$ to a maximum value
(given in table~\ref{tb:droplets}) when the vortex becomes unstable to
ripplon formation.  For a molecule with a translational mass of
100\,u, the maximum kinetic energy that the molecule requires to stay
bound to a curved vortex line is between 19.1 and 1.1\,K as $N$ varies
between $10^2$ and $10^6$ (see Table~\ref{tb:droplets}).  This kinetic
energy can be compared to binding energies to the straight vortex of
5.0, 4.4, and 7.7\,K for Xe, HCN, and SF$_6$, respectively, as
calculated by DFT.\cite{Dalfovo00}  Thus, for the least stable curved
vortex lines in smaller droplets, the molecules may become unpinned.
This suggests that it would be useful to have improved binding energy
estimates for molecules, particularly to curved vortex states.

\section{Vortex Formation Mechanisms}

A possible explanation for the failure of existing experiments to
detect droplets with vortex lines is that they cannot be formed in the
first place, perhaps because of the high energy and angular momentum
required.  We will consider the probability of events that deposit
sufficient energy and angular momentum in the droplets such that
vortex formation can in principle take place.

In most experiments, the droplets grow out of a supercooled
gas.\cite{Northby01} In this situation, the large droplets probably
grow at least in part by coalescence of smaller droplets.  If we
consider the coalescence of two droplets with $N/2$ atoms that collide
with an average impact parameter (2/3 of either's diameter), the
minimum relative velocity required to deposit an angular momentum of
$N \hbar$ in the combined droplet is $v\si{rel}=270 N^{-1/3}$\,m/s,
which is $58-2.7$\,m/s for $N=10^2-10^6$.  Experiments of Toennies
\emph{et al.}\cite{Buchenau90} have found that the expansions produce
droplets with a final speed ratio of $\approx 100$, which implies
final relative velocities of $\approx 8$\,m/s for expansion from a
20\,K source.  While such relative collision velocities are, except
for the largest droplets, less than those required to produce a vortex
line with maximum angular momentum (straight vortex), they are
sufficient to produce curved vortices.  Further, the above estimates
are only average values, and the relative velocities in the part of
the expansion where the droplets undergo substantial growth are likely
significantly higher.  Experiments that produce droplets above
$N\approx \pt{5}{4}$ typically use expansion conditions such that
liquid helium is ejected through a cold nozzle into vacuum, and
fragments by cavitation as the pressure falls far below the
equilibrium vapor pressure.  This break-up is believed to generate
considerable turbulence, and thus may lead to dense vortex
formation.\cite{Zurek85,Dodd99}

The ``pick-up'' process,\cite{Gough85} by which droplets are doped
with solutes, should often deposit enough angular momentum, and almost
always enough energy, to form a linear vortex line, particularly for
the pickup of heavy molecules by not too large droplets.  It has been
predicted that vortices are nucleated when the velocity of an impurity
exceeds to sound velocity in helium,\cite{Berloff00} which is well
below at least the typical impact speed of atoms or molecules striking
helium nanodroplets.  Consider the pickup of atoms or molecules of
mass $M$ from a thermal gas at temperature $T$ by droplets with
laboratory frame velocity $v\si{d}$.  Integration over the
Maxwell--Boltzmann distribution for the gas gives the probability
density $P\si{r}$ for the relative velocity $v\si{r}$ between droplet
and gas
\begin{equation}
P\si{r}(v\si{r}) = \sqrt{\frac{2M}{\pi k T}}
\frac{v\si{r}}{v\si{d}}
\sinh \left[ \frac{M v\si{r} v\si{d}}{k T} \right]
\exp \left[ -\frac{M(v\si{r}^2+v\si{d}^2)}{2 k T} \right].
\end{equation}
From this distribution, the average relative collisional speed,
$\bar{v}\si{r}$, is calculated to be
\begin{equation}
\label{eq:vbar}
\bar{v}\si{r} = \sqrt{\frac{2 k T}{\pi M}} \exp \left[
-\frac{M v\si{d}^2}{2 k T} \right]
+ \left(v\si{d} + \frac{k T}{M v\si{d}} \right) 
\mathrm{erf}\sqrt{\frac{M v\si{d}^2}{2 k T}}.
\end{equation}

For a given impact parameter $b$, all collisions with relative
velocity $v\si{r}\ge v\si{m}(b)=\hbar/b \cdot (1/m+N/M)$ result in
angular momentum of at least $N \hbar$, enough to form a straight
vortex. The fraction of such collisions is
\begin{equation}
\Phi(b) = \frac{\int_{v_{\textrm{\tiny{m}}}(b)}^{\infty}
v\si{r} P\si{r}(v\si{r})
\mathrm{d}v\si{r}}
{\int_0^{\infty} v\si{r} P\si{r}(v\si{r})
\mathrm{d}v\si{r}}.
\end{equation}
If we assume that the the pickup probability is independent of
collisional impact parameter $b$ for $b \le R = r_0 N^{1/3}$, and
falls abruptly to zero for $b > R$, then the fraction of resulting
doped droplets that have at least $N \hbar$ of angular momentum from
the pickup process is
\begin{equation}
P_{\ge N \hbar} = \int_0^R \frac{2\pi b\, \mathrm{d}b}{\pi R^2} \Phi(b).
\end{equation}

\begin{figure}[htbp]
\includegraphics[width=3in]{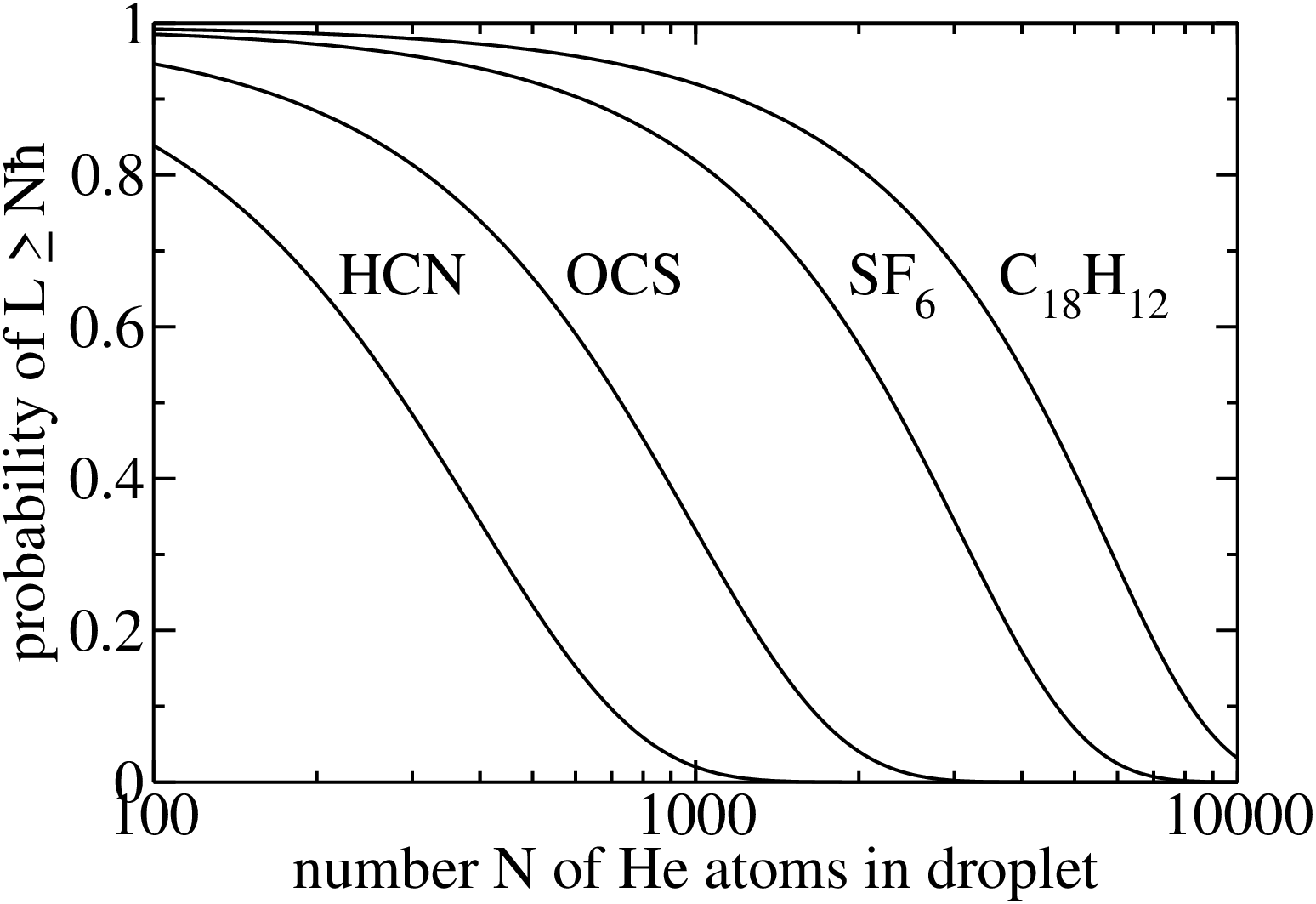}
  \caption{Probability of a pickup collision depositing
    more than $N \hbar$ units of angular momentum in the droplet, as
    function of droplet size $N$. Droplet velocity $v\si{d}=456$\,m/s
    (20\,K supersonic expansion); gas temperature $T=300$\,K.}
  \label{fig:collisions}
\end{figure}
\begin{figure}[htbp]
\includegraphics[width=3in]{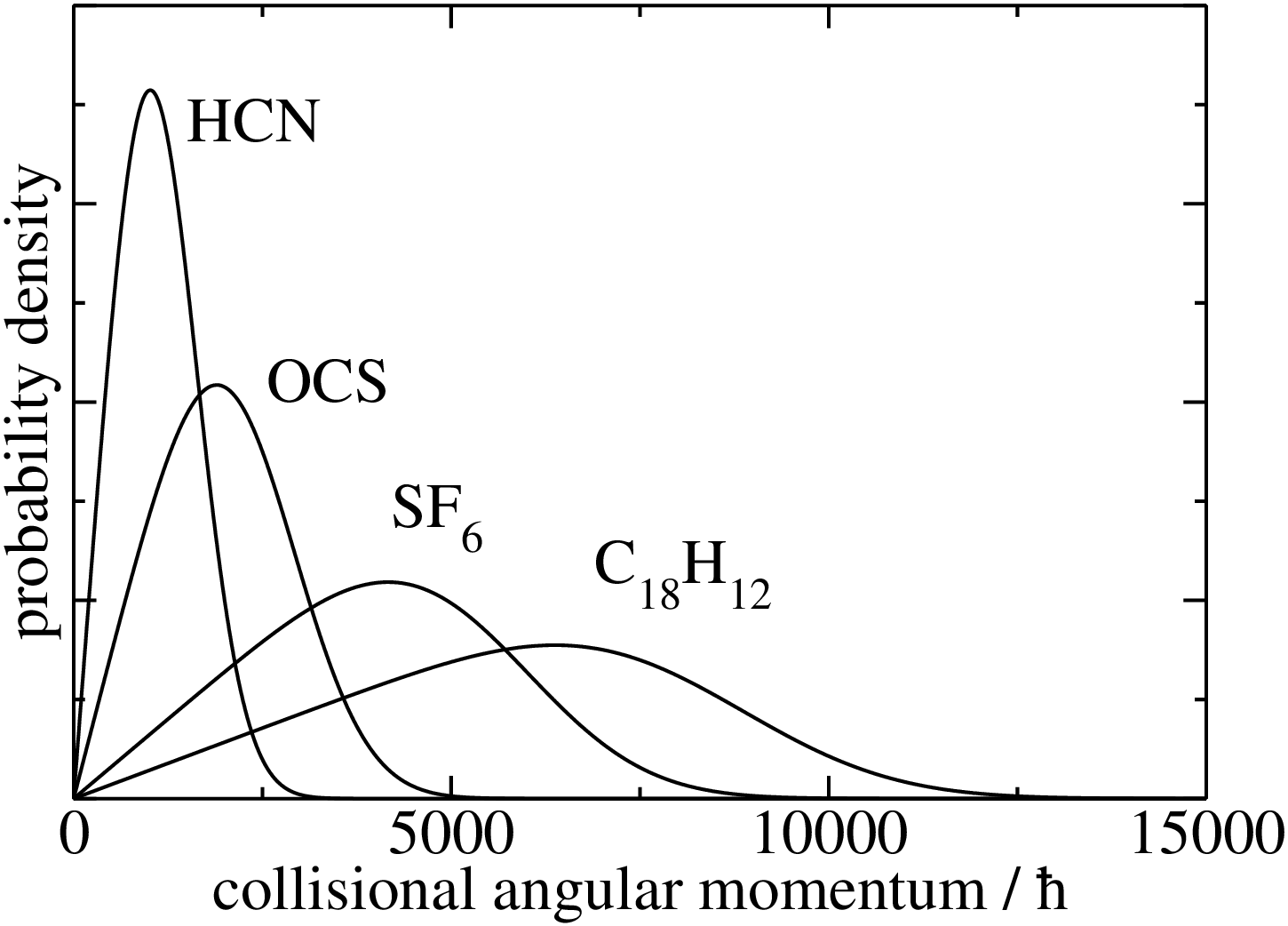}
  \caption{Probability densities of collisional
    angular momenta during pickup of various molecules by a droplet
    with $N = 10^4$ He atoms. Droplet velocity $v\si{d}=456$\,m/s
    (20\,K supersonic expansion); gas temperature $T=300$\,K. For
    other droplet sizes, the shapes of the distributions are the same
    except for a correction due to changes in the reduced mass, and
    scaling of the average collisional angular momentum as $N^{1/3}$.}
  \label{fig:L_dist}
\end{figure}

Figure~\ref{fig:collisions} shows plots of the fraction of pickup
collisions that result in an angular momentum greater than $N \hbar$,
as a function of $N$ for a number of solutes.  Figure~\ref{fig:L_dist}
shows, for droplets of $N = 10^4$ helium atoms, the distribution of
collisional angular momenta deposited by pickup of the same solutes.
In many cases relevant to previously reported experiments, the pickup
process has a fair probability to deposit sufficient angular momentum
to create a straight vortex, and almost always sufficient angular
momentum to create curved vortex lines.  Thus, the lack of observation
of vortex lines does not appear to be due to a lack of initial angular
momentum to form them.

\section{Stability Analysis of Vortex States}

An alternative explanation for the lack of evidence of vortex lines in
experiments to date could be that vortex lines are unstable, and decay
or are expelled from the droplets in a time short compared to the time
between the pickup of solutes and the spectroscopic detection of the
doped droplets.  In this section, several possible vortex destruction
pathways are considered, concentrating on the simplest case of the
linear vortex line.

A helium nanodroplet excited with a linear vortex line has an angular
momentum of $N \hbar$, much higher than the magnitude of the angular
momentum thermally present in the ripplons at the droplet temperature
of $\approx 0.38$\,K that has been experimentally found for helium
nanodroplets cooled by evaporation.\cite{Hartmann95}  For droplets in
the size range of experiments reported to date, the thermal excitation
of phonon states is negligible in comparison to that of the
ripplons.\cite{Brink90}  The lowest frequency ($L = 2$) ripplon mode
has the lowest energy per unit angular momentum of any of the helium
excitations except vortices, but this value is still higher than the
corresponding value for the linear vortex, as indicated in
Figure~\ref{fig:Esv}; Table~\ref{tb:droplets} shows the ratio of
excitation energies and angular momenta for the $L=2$ ripplon and the
linear vortex for a range of helium droplet sizes.  While decay of the
vortex state is highly exothermic, this is only true if one ignores
the requirement of conservation of angular momentum; this neglect is
appropriate for helium contained in a vessel with walls, to which the
angular momentum can be transferred, but not for a droplet isolated in
a high vacuum chamber.  When angular momentum conservation is
enforced, the decay of the linear vortex line into ripplons is highly
endothermic and thus cannot occur for a droplet unless the droplet
also contains a counter-angular momentum in ripplons that is nearly as
large as the angular momentum of the vortex itself.
Table~\ref{tb:droplets} also contains the thermal average value
$\sqrt{\left< L(L+1) \right>}$ of ripplon angular momentum for
different size droplets at 0.38\,K:\cite{Lehmann99b} it is only a
small fraction of that of the straight vortex.  Exchange of angular
momentum between thermally populated ripplons and a vortex could at
most allow the vortex to move slightly off axis.

Helium nanodroplets can shed angular momentum by helium atom
evaporation.  A helium atom evaporating with momentum $p$ can carry
away a maximum angular momentum of $\approx p R$, which reduces the
internal energy of the droplet by $E\si{b} + p^2/(2m\si{He})$, where
$E\si{b}$ is the binding energy of the helium atoms to the droplet.
For finite droplets, $E\si{b}$ is reduced from the bulk value due to a
term that reflects the change in surface energy upon evaporation of an
atom.\footnote{ By differentiation of the sum of bulk and surface
  energies with respect to He particle number, one finds that
  $E\ssi{b}(N) = E\ssi{b}(\textrm{bulk}) - \frac{8\pi}{3} \sigma r_0^2
  N^{-1/3}$.}  Table~\ref{tb:droplets} shows the maximum angular
momentum that can be lost by single atom evaporation using the entire
energy of the linear vortex line.\footnote{This angular momentum is
  given by $\sqrt{2 m\ssi{He} (E\ssi{HC}(N) - E\ssi{b}(N))} \cdot
  R(N)$.}  In every case, that angular momentum is far lower than the
angular momentum of the vortex line that must be carried off for the
vortex line to decay.  Evaporation of multiple helium atoms will
increase the angular momentum that can be carried off for the same
total helium kinetic energy.  Because the available kinetic energy is
reduced by the increased binding energy of the multiple helium atoms,
there is a number of evaporated atoms that allows the maximum angular
momentum to be removed.  Table~\ref{tb:droplets} also lists this
amount of angular momentum for a range of droplet sizes.\footnote{This
  maximum angular momentum is the maximum of $\sqrt{2m\ssi{He}(n
    E\ssi{HC}(N) - n^2 E\ssi{b}(N))} \cdot R(N)$ with respect to $n$,
  the number of evaporated helium atoms.}  It is evident that in all
cases, this maximal amount of angular momentum lost by evaporation is
far below that which is needed for the vortex line to decay while
conserving angular momentum.

The above analysis assumed that the helium atoms evaporated as
isolated atoms.  An alternative decay mechanism is fission of the
droplet: in this case, the loss of helium binding energy is only due
to the increase in surface energy, which is higher for the two
droplets than for the original nearly-spherical droplet.  Fission of a
droplet into two equal-size droplets costs far more energy than that
contained in a vortex line, even neglecting the required kinetic
energy for relative motion of the fragments.  Table~\ref{tb:droplets}
contains the fragmentation that carries away the maximum possible
quantity of angular momentum, taking the original droplet radius as
the impact parameter of the departing fragments.\footnote{This angular
  momentum is the maximum with respect to the number of atoms in one
  fragment, $n$, of $\sqrt{2 \mu(n) (E\ssi{HC}(N) - \Delta S(n,N)
    \sigma )} \cdot R(N)$ where $\mu(n) = m\ssi{He} n(N-n)/N$ is the
  reduced mass of the departing fragments and $\Delta S = 4 \pi r_0^2
  (n^{2/3} + (N-n)^{2/3} - N^{2/3})$ is the increase in surface area.}
Yet again, this is far below the angular momentum of the linear
vortex.

The above considerations show that the simultaneous constraints of
conservation of energy and angular momentum prevent decay of straight
vortices in a pure droplet by any of the mechanisms that are known to
the authors.  Adding a molecule to the droplet will further decrease
the energy per unit angular momentum of the vortex as long as the
molecule remains bound to the vortex.  It thus appears that droplets
with a vortex line should be stable to decay.

How can we understand the failure to date to observe vortex lines in
helium nanodroplet isolation spectroscopy?  The answer is probably
related to the fact that under typical experimental conditions, the
above mechanisms, by putting sufficient angular momentum into the
droplets to form vortex lines, add far more than the minimum energy
required to do so.  The density of ripplon states grows very rapidly
with increasing energy;\cite{Lehmann_in_press} it may well be that at
the energy and angular momentum of a droplet following coalescence or
pickup, the fraction of states that contain a vortex line is a totally
insignificant fraction of the total density of states, and that
droplets will shed sufficient angular momentum along with energy as
they evaporatively cool that they have a negligible chance to be
trapped in one of the metastable solutions containing a vortex.  We
have recently carried out Monte Carlo cooling calculations, using
statistical reaction rate theory, for the evaporative cooling of pure
and doped helium nanodroplets, conserving both energy and angular
momentum,\cite{Lehmann_Dokter_up} considering only ripplon
excitations of the droplets.  In the future, we hope to extend that
work to include the spectrum of vortex line excitations as well, which
it is hoped will shed further light on the still unsolved problem of
why spectra of molecules bound to vortex lines has yet to be observed
in Helium Nanodroplet Isolation Spectroscopy.

\appendix

\section{Hollow-Core Model}
\label{sec:HCmodel}

In the local-induction approximation, the flow field around a curved
vortex is approximated locally by the flow around a vortex ring of
equal radius of curvature. The velocity potential for the flow around
a vortex ring in the $x-y$ plane, with radius $\mathcal{R}$, centered at the
origin and moving in the $+z$ direction, can be computed from
Eq.~(\ref{eq:BS}):
\begin{widetext}
\begin{multline}
\label{eq:phiexact}
\phi = \frac{\kappa \mathcal{R} z}{\pi(\mathcal{R}^2-r^2-z^2)\sqrt{(\mathcal{R}+r)^2+z^2}}
\left( (\mathcal{R}+\sqrt{r^2+z^2}) \cdot
\Xi \left[ \frac{(\mathcal{R}-r)^2+z^2}{\mathcal{R}^2+r^2+z^2-2\mathcal{R}\sqrt{r^2+z^2}},
\frac{(\mathcal{R}-r)^2+z^2}{(\mathcal{R}+r)^2+z^2} \right]\right.\\ +
\left.(\mathcal{R}-\sqrt{r^2+z^2}) \cdot
\Xi \left[ \frac{(\mathcal{R}-r)^2+z^2}{\mathcal{R}^2+r^2+z^2+2\mathcal{R}\sqrt{r^2+z^2}},
\frac{(\mathcal{R}-r)^2+z^2}{(\mathcal{R}+r)^2+z^2} \right] \right)
+ \frac{\kappa}{2} \textrm{sign}(z),
\end{multline}
\end{widetext}
with the function
\begin{equation}
\Xi[n,m] = i \left( \Pi(n|m) - \frac{1}{\sqrt{m}}
\Pi(\frac{n}{m} | \frac{1}{m}) \right)
\end{equation}
in terms of complete elliptic integrals of the third kind, $\Pi(n|m)$;
$\kappa=h/m$ is the quantum of circulation.  Series expansion of
Eq.~(\ref{eq:phiexact}) around the vortex singularity, with $(r,z)=\mathcal{R}
\cdot (1+\zeta \cos \vartheta, \zeta \sin \vartheta)$ and $0<\zeta\ll
1$, yields
\begin{equation}
\label{eq:phiapprox}
\phi \approx \frac{\kappa \vartheta}{2\pi} - 
\frac{\kappa \zeta}{4\pi} \ln \frac{8}{\zeta} \sin \vartheta
+ \frac{3\kappa \zeta^2}{32\pi} (\ln \frac{8}{\zeta}
- \frac56 ) \sin 2\vartheta.
\end{equation}
In the hollow-core model, we assume that the vortex singularity is
surrounded by an empty ``vortex core'' of radius $a$, outside of which
the helium density is constant and equal to the bulk value, \density.
The helium velocity field is given by $\vect{v} = -\vect{\nabla}\phi$:
the first term in Eq.~(\ref{eq:phiapprox}) is the source of the
looping helium flow around the vortex, the second term gives rise to
the forward motion of the vortex ring with speed
\begin{equation}
\label{eq:forward}
u=\frac{\kappa}{4\pi \mathcal{R}} \left[ \ln \frac{8\mathcal{R}}{a}-1 \right]
\end{equation}
along the $+z$ axis, and the third term allows us to estimate the
residual flow across the vortex core surface due to the fact that the
exact vortex core surface is not circular in cross section: the ratio
of the rms velocity of this residual normal flow to the forward
velocity is $3a/(4\mathcal{R}\sqrt{2})$ as $a/\mathcal{R}\to 0$.

Rayfield and Reif\cite{Rayfield64} have experimentally determined the
forward velocity $u$ of ring vortices, and determined a core radius of
$\tilde{a}=1.00$\,\AA\ by using the formula $u = \frac{\kappa}{4\pi \mathcal{R}}
\left[ \ln \frac{8\mathcal{R}}{\tilde{a}}-\frac12 \right]$. Comparison of this
expression to Eq.~(\ref{eq:forward}) yields $a = \tilde{a}/\sqrt{e} =
0.607$\,\AA, which is the core radius that we use in this work for the
hollow-core model.

\section{Vortex Shape Calculation}
\label{sec:shape}

A vortex, $\vect{s}(\len)$, is most generally described as a curve in
the $x, z$ plane parametrized by its length $\len$; three-dimensional
vortex curves are always longer and thus higher in energy for given
boundary conditions.\cite{Bauer95}  The local vortex curvature is
\begin{equation}
\label{eq:curvature}
\chi(\len) = \frac{x''(\len)}{z'(\len)} = -\frac{z''(\len)}{x'(\len)}.
\end{equation}
In the hollow-core model with the local-induction approximation, the
magnitude of the vortex velocity is determined from the core radius
and the local curvature, and its direction is given by the binormal
vector (the normalized cross product of $\frac{d\vect{s}}{d\len}$ and
$\frac{d^2\vect{s}}{d\len^2}$).  The magnitude of the velocity is
given by Eq.~(\ref{eq:forward}):
\begin{equation}
\label{eq:implicitshape}
v(\len) = -\frac{\kappa}{4\pi} \chi(\len) \left[
\ln \frac{a \chi(\len)}{8}+1 \right].
\end{equation}
For a vortex to rotate around the $z$ axis at angular velocity
$\Omega$ without changing shape, it must satisfy $v(\len) = \Omega
x(\len)$, which leads to the coupled differential equations that
determine the vortex shape:
\begin{equation}
\label{eq:explicitshape}
\{ x''(\len),z''(\len) \} =
\frac{\omega x(\len)}{W_{-1}\left[-a \omega x(\len)e/8\right]}
\{ -z'(\len),x'(\len) \},
\end{equation}
where we have used $\omega=4\pi\Omega/\kappa$ and the Lambert function
$W_{-1}(x)$\footnote{In \emph{Maple}, $W_k(x)$ is entered as
  \texttt{LambertW(k,x)}; in \emph{Mathematica},
  \texttt{ProductLog[k,x]}. An efficient numerical routine for the
  evaluation of the Lambert function is given in
  Ref.~\onlinecite{Chapeau-Blondeau02}.} defined as the smaller of the two
real solutions of $x=y e^y$ for $-1/e\le x<0$.  Combined with the
initial conditions $\{x(0),z(0)\}=\{x_0,0\}$ and
$\{x'(0),z'(0)\}=\{0,1\}$, Eq.~(\ref{eq:explicitshape}) yields the
desired vortex shape functions by numerical integration. For a given
minimum approach distance $x_0$ to the $z$ axis, we must pick the
parameter $\omega$, and thus the angular velocity $\Omega$, such that
the resulting vortex line intersects the droplet surface
perpendicularly; for a spherical droplet with radius $R$, this
condition is $x(\len_1)z'(\len_1)=z(\len_1)x'(\len_1)$ with
$x(\len_1)^2+z(\len_1)^2=R^2$.

This procedure differs from the one given in Ref.~\onlinecite{Bauer95} in
that Bauer \emph{et al.}\ neglect the variation of the Lambert
function, and replace the denominator of Eq.~(\ref{eq:explicitshape})
by a constant. For droplets consisting of $100\le N \le 100000$ helium
atoms, the error of this approximation is below 0.6\% for the
evaluation of the angular velocity.

Figure~\ref{fig:Omega} shows angular velocities of curved vortices
computed with both models.

As $x_0\to R$, the vortex solutions are close to half-circles with
radius $R-x_0$. We expect the hollow-core model to break down if
$R-x_0 \lesssim a$.  A clear sign of this breakdown is that for
$x_0\to R$, the Lambert function in Eq.~(\ref{eq:explicitshape})
starts giving complex values as the droplet surface is approached ($a
\omega x(\len) e/8 > 1/e$).  In the simplified model of
Ref.~\onlinecite{Bauer95}, the angular velocity turns toward negative values
for $R-x_0<a$, invalidating that model as well.  It is to be noted
that in this limit, the description of the vortex core as a
cylindrical tube around the vortex is very inaccurate.

\section{Core Corrections}

Bauer \emph{et al.}\ show\cite{Bauer95} that the angular momentum and
kinetic energy of a droplet with vortex can be written as surface
integrals:
\begin{align}
\label{eq:surfL}
\vect{L}\si{v} & = m \density \int_{\Sigma} \phi \vect{r} \times
\mathrm{d}\vect{S}\\
\label{eq:surfE}
E\si{v} & = \frac{1}{2} m \density \int_{\Sigma} \phi \vect{v} \cdot
\mathrm{d}\vect{S},
\end{align}
where $\Sigma$ is the surface of a connected volume of helium with no
branch cuts in the velocity potential $\phi$, and
$\vect{v}=-\vect{\nabla}\phi$. While most of the angular momentum and
kinetic energy comes from the surface integrations of
Eqs.~(\ref{eq:L},\ref{eq:Ev}), there are corrections due to the
surface integrations of Eqs.~(\ref{eq:surfL},\ref{eq:surfE}) over the
vortex core. Using Eqs.~(\ref{eq:phiapprox}), (\ref{eq:curvature}),
and (\ref{eq:explicitshape}), we find these corrections to be
\begin{align}
  \delta L_{\textrm{\scriptsize{v}},z} & \approx \frac{m \rho
    \kappa}{a \omega^2} \left[ C_{1,0} - \frac14 C_{2,0}+
    \frac12 C_{2,1}\right]\\
  \delta E\si{v} & \approx \frac{m \rho \kappa^2}{8\pi a \omega}
  \left[ -C_{1,0} + \frac14 C_{2,0} - \frac18 C_{2,1}
\right],
\end{align}
with
\begin{equation}
C_{n,m}=(a \omega)^{n+1}\int\si{vortex} \frac{x(\len)^n}
{W_{-1}^m(-a \omega x(\len) e/8)} \mathrm{d}\len.
\end{equation}
From symmetry considerations, we must have
$L_{\textrm{\scriptsize{v}},z}(-x_0) =
L_{\textrm{\scriptsize{v}},z}(x_0)$ and $E\si{v}(-x_0) =
E\si{v}(x_0)$. In fact, the first core correction ($C_{1,0}$) adjusts
$L_{\textrm{\scriptsize{v}},z}$ to satisfy this symmetry constraint,
\emph{i.e.}, $L_{\textrm{\scriptsize{v}},z}(x_0) = N \hbar -
\mathcal{O}(x_0^2)$. However, this symmetry is not exactly satisfied
for $E\si{v}(x_0)$, which we assume is due to inaccuracies of the
local-induction approximation.

\begin{acknowledgments}
  
  The authors would like to thank Manuel Barranco for helpful comments
  upon an draft of this paper.  This work was supported by a grant
  from the National Science Foundation.

\end{acknowledgments}

\bibliography{vortex}

\end{document}